\documentclass[prl,twocolumn,showpacs,preprintnumbers,amsmath,amssymb]{revtex4}
\usepackage{graphicx}

\begin{document}

\title{Structural relaxations in electronically excited
poly(\textit{para}-phenylene)}

\author{Emilio~Artacho$^1$, M.~Rohlfing$^2$, M.~C\^ot\'e$^3$, 
P.~D.~Haynes$^4$, R.~J.~Needs$^4$, and C.~Molteni$^5$}
\affiliation{
$^1$Department of Earth Sciences, University of Cambridge,
      Downing Street, Cambridge, CB2 3EQ, UK\\
$^2$School of Engineering and Science,
      International University Bremen, P.O.Box 750561,
      28725 Bremen, Germany\\
$^3$D\'epartement de Physique, Universit\'e de Montr\'eal, C.P.\ 6128, 
      succ.\ Centre-Ville, Montreal, Quebec, H3C 3J7, Canada\\
$^4$TCM Group, Cavendish Laboratory, University of Cambridge,
      Madingley Road, Cambridge, CB3 0HE, UK\\
$^5$Physics Department, King's College London, Strand, London WC2R 2LS, UK
}

\date{6 Februrary 2004}

\begin{abstract}
Structural relaxations in electronically excited
poly(\textit{para}-phenylene) are studied using many-body perturbation
theory and density-functional-theory methods.  A sophisticated
description of the electron-hole interaction is required to describe
the energies of the excitonic states, but we show that the structural
relaxations associated with exciton formation can be obtained quite
accurately within a constrained density-functional-theory approach.
We find that the structural relaxations in the low-energy excitonic
states extend over about 8 monomers, leading to an energy reduction of
0.22 eV and a Stokes shift of 0.40 eV.
\end{abstract}


\maketitle

The sustained growth of interest in conjugated polymers arises from
their potential as the active material in low-cost field-effect
transistors, photovoltaic devices and light emitting diodes
(LEDs).\cite{greenham_1995,heeger_2001} Investigating electronically
excited states is an important part of this effort, which has proved a
great challenge for theoretical techniques.  An accurate description
of the excited electronic states of conjugated polymers requires the
inclusion of the strong electron-hole interaction and the structural
relaxations which occur in the excited state.

In its simplest form, a polymer LED consists of a layer of conjugated
polymer sandwiched between two electrodes.  Electrons and holes are
injected into the polymer where they are attracted to one another and
form bound excitons. Light is then emitted by exciton recombination.
In an excited state the polymer may lower its energy by structural
relaxations so that the peak optical emission frequency is lower than
the peak absorption frequency.  Although this ``Stokes shift'' can, at
least in principle, be measured, the actual relaxations have not been
determined experimentally.  These relaxations localize excitons and
are relevant for the technologically important processes of exciton
migration and recombination.

We have investigated structural relaxations in the low-energy excited
states of poly(\textit{para}-phenylene) (PPP).  A PPP LED has been
demonstrated which emits blue light in a band around
2.7~eV.\cite{grem_1992,grem_1993} PPP is, however, insoluble and
therefore difficult to process and, instead, soluble derivatives of
PPP with various side groups are preferred for manufacturing LEDs.
The low-energy optical properties of PPP and its derivatives are
similar, although the structural relaxations in the excited states
depend on the nature of the side groups.  We have chosen to study PPP
because it has a simpler structure than its derivatives.

\begin{figure}[ht]
\centering\includegraphics[width=7cm]{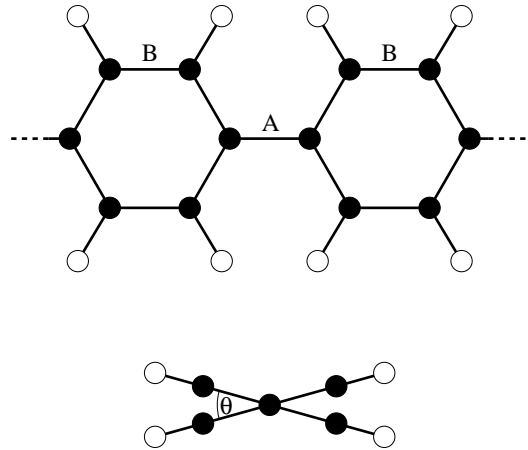}
\caption{The structure of a PPP chain with the C atoms represented by
filled circles and the H atoms by open circles.  (a) Side view showing
the A and B bonds.  (b) End view showing the torsional angle
$\theta$.}
\label{fig:structure}
\end{figure}

As we show below, systems containing more than 100 atoms are required
for studying the excited-state relaxations of PPP, which pose a
significant problem to theory.  Such systems can easily be treated by
density-functional theory (DFT) \cite{footnote_lda}, but DFT alone
often has difficulties in describing excited states.  Excited states
can be described by many-body perturbation theory (MBPT) \cite{bse},
but this is numerically demanding and cannot currently be applied to
such large systems. However, we find that the forces in the excited
states of PPP obtained within DFT are very similar to those within
MBPT, which indicates that the relaxed structures and Stokes shift
obtained within DFT are reliable.

A PPP chain (Fig.~\ref{fig:structure}) consists of phenyl rings joined
by single C-C bonds, with adjacent rings being rotated with respect to
one another by a torsional angle $\theta$.  Our DFT structures for the
ground states of the single chain and crystal are in excellent
agreement with a previous DFT calculation and in accord with
experiment.\cite{ambrosch-draxl_1995} Our DFT band gap for the single
chain is 2.44 eV and that for the crystal is 1.70 eV, which is in good
agreement with a previous DFT calculation.\cite{ambrosch-draxl_1995}

We began by using a simple model of how an optical excitation affects
the bonding in PPP: we performed DFT calculations of the changes in
the electronic charge distribution due to a single excitation from the
highest occupied molecular orbital (HOMO) to the lowest unoccupied
molecular orbital (LUMO). In both the isolated chain and crystalline
forms we found an increase in the electronic charge in the A bonds and
to a lesser extent in the B bonds (see Fig.~\ref{fig:structure}).
These changes indicate the development of double-bonding character
which is expected to favor both a more planar arrangement of rings
(reduction in $\theta$) and compression of the A and B bonds.

In a perfect chain or crystal an exciton would, in principle, be
delocalized because localization within some region of space costs
kinetic energy.  However, a larger energy may be gained by structural
relaxations so that an exciton may localize itself.  The low-energy
excitonic states of PPP may therefore be self-localized by structural
relaxations, i.e., local reductions in $\theta$ and the lengths of the
A and B bonds.

\begin{figure}[ht]
\centering\includegraphics[width=8cm]{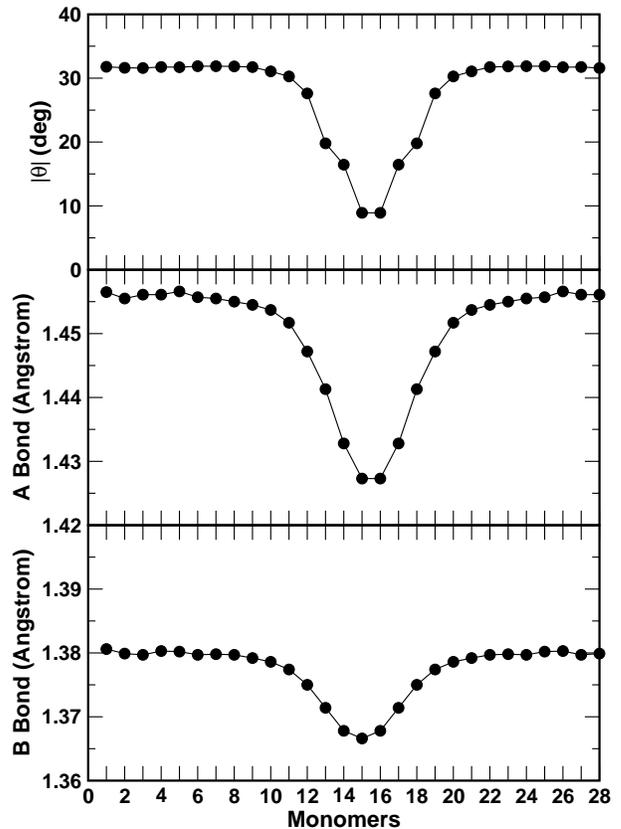}
\caption{The torsional angle, $\theta$, and the lengths of the A and B
bonds (see Fig.~\ref{fig:structure}) as a function of monomer number
for the excited state of a single PPP chain.}
\label{fig:bond+torsion}
\end{figure}

The structural relaxations in the lowest excited states of a single
PPP chain can be investigated using a constrained DFT method.  To this
end we model the excited state by promoting an electron from the HOMO
to the LUMO at wave vector $k=0$ and performing full structural
relaxations in unit cells of various sizes.  A unit cell containing 28
monomers was found to be sufficient to converge the excited state
structure.  The important parameters of the minimum energy structure
are shown in Fig.~\ref{fig:bond+torsion}.  The relaxations are largely
confined to a region of approximately 8 monomers over which the
torsional angle and the lengths of the A bonds are significantly
reduced, and the lengths of the B bonds are somewhat reduced.  The
gain in energy from the relaxation is 0.22 eV, and the Stokes shift,
calculated as the difference between the band gaps of the ground and
excited state structures, is 0.40 eV.\cite{footnote_gaps}

We have analyzed the localization of the exciton in terms of the
compression of the A and B bonds and the reduction in the torsional
angle, $\theta$.  We performed two constrained DFT calculations for
the excited state in which (1) only relaxations along the axis of the
chain and (2) only relaxations in the plane perpendicular to the chain
were allowed.  The relaxation energies of 0.14 eV and 0.17 eV,
respectively, add up to significantly more than the full relaxation
energy of 0.22 eV, indicating that the modes are strongly coupled.
The relaxations along the chain led to a Stokes shift of only 0.06 eV
and to very weak localization of the exciton, whereas the relaxations
perpendicular to the chain gave a Stokes shift of 0.16 eV and much
stronger localization.  The torsional relaxation is therefore very
important in determining the Stokes shift, although the coupling to
the bond compression is strong and both types of relaxation are
important in determining the full relaxation energy and Stokes shift.

Stokes shifts in PPP derivatives have been measured in dilute
solutions and solid films.  A common approach is to use ``laddered''
PPP derivatives in which some or all of the phenyl rings are joined to
their neighbors by chemical ``bridges''.  These bridges hinder
rotation about the A bonds which, according to our picture, is
expected to reduce the Stokes shift.  Hertel \textit{et
al.}~\cite{hertel_2001} report absorption and emission measurements on
a series of laddered and non-laddered polymers, which clearly show
that laddering reduces the Stokes shift. As shown in Fig.~3b of Hertel
\textit{et al.}~\cite{hertel_2001}, in dilute solution the
non-laddered polymer dodecyloxy-poly(\textit{para}-phenylene)
(DDO-PPP) exhibits a Stokes shift of 0.6$\pm 0.2$ eV.  Considering the
presence of the solvent and different side chains in the experimental
system the agreement with our single chain PPP value of 0.40 eV is
reasonable.  The measured Stokes shifts for the laddered polymers are
smaller, in agreement with our picture that changes in $\theta$ couple
strongly to the band gap.  Our picture also suggests that structural
relaxations for excitons in laddered PPP derivatives should only lead
to a weak localization effect.

We now investigate the reliability of the constrained-DFT results for
the relaxation and Stokes shift of the exciton.  The formation of a
completely delocalized exciton leads to a vanishingly small change in
the charge density and therefore vanishingly small forces on the
atoms.  If, however, a lattice distortion occurs which reduces the
band gap in some region of space then the exciton will be attracted to
this region.  If the lowering of the energy of the exciton is greater
than the energy required to form the distortion then the exciton will
stabilize the distortion, leading to a self-localized exciton.  The
reduction in the band gap due to a lattice distortion is an effect
which is described approximately within our DFT calculations.  The
excitonic energies themselves, however, are strongly modified by
many-body effects arising from the electron-hole interaction.  We now
investigate the dependence of these many-body effects on the lattice
distortion.  If this dependence is sufficiently weak the constrained
DFT approach will be reliable.

A rigorous approach to electronic many-body effects for excited
electronic states is given by many-body perturbation theory
(MBPT).\cite{bse} We have used the $GW$ approximation to describe the
addition or removal of an electron and the Bethe-Salpeter equation
(BSE) for the excitation of an electron including the electron-hole
interaction.  These techniques have recently been used to describe the
excitonic states of solids~\cite{rohlfing_2000},
clusters~\cite{rohlfing_1998}, and polymers.\cite{rohlfing_1999,horst}
MBPT is numerically very demanding and cannot currently be applied
directly to a polymer with a self-localized exciton.

\begin{figure}[ht]
\centering\includegraphics[width=8cm]{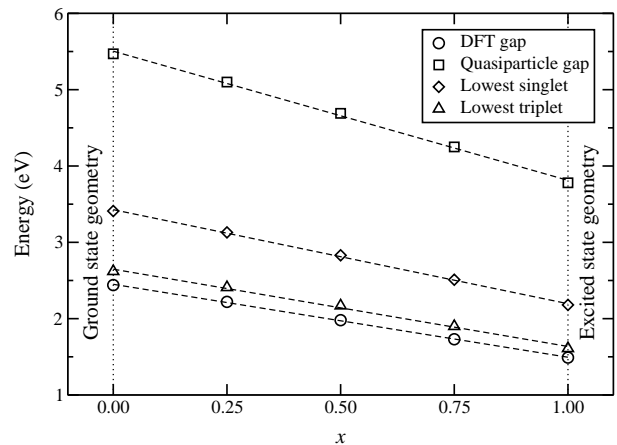}
\caption{Excitation energies calculated within the DFT and $GW$
approaches and the $GW$-BSE singlet and triplet energies as a function
of the structural parameter $x$ for a single PPP chain.}
\label{fig:energy_x}
\end{figure}

To investigate the effects of electronic correlation and validate the
constrained-DFT approach we proceeded as follows.  We performed DFT
and MBPT calculations for a series of structures involving torsional
angles and compressions of the A and B bonds which are similar to
those found in the self-localized exciton of
Fig.~\ref{fig:bond+torsion}.  We considered the relaxation of the two
central monomers of Fig.~\ref{fig:bond+torsion} as indicating the
maximum relaxations which occur in the exciton.  We then periodically
repeated this structure, obtaining a reference system which can be
investigated within both constrained DFT and MBPT.  We defined a
structural parameter $x$ which takes the value $x$=0 for the ground
state structure ($\theta$=33.7$^{\circ}$, A=1.456 \AA, and B=1.380
\AA), and $x$=1 for the geometry corresponding to the central monomers
of the self-localized exciton ($\theta$=9.1$^{\circ}$, A=1.427 \AA,
and B=1.366 \AA).  Intermediate values of $x$ correspond to linearly
interpolating $\theta$, A and B between the extremal values.
Fig.~\ref{fig:energy_x} shows the excitation energies of this periodic
structure as a function of $x$.  The four curves denote the DFT energy
gap, the quasiparticle (QP) energy gap, and the transition energies of
the lowest spin-singlet and spin-triplet excitons obtained from
solving the BSE.\cite{note_bse_technical} The QP gaps are 2.3-3.0 eV
larger than the DFT gaps due to the significant QP corrections typical
of semiconducting systems.  The exciton energies, on the other hand,
are 1.5-2.9 eV smaller than the QP gaps due to the attractive
electron-hole interaction.  The singlet (triplet) excitation energy
ranges from 3.41 eV (2.62 eV) at $x$=0 to 2.18 eV (1.61 eV) at $x$=1.

The change in the excitation energy when going from $x$=0 to $x$=1
constitutes the excited-state contribution to the force on the
structure projected onto the relaxations studied.  The most important
feature of Fig.~\ref{fig:energy_x} is that this change (--1.22 eV or
-1.01 eV, respectively, for the singlet and triplet state) is very
similar to the change in the DFT gap energy (--0.95 eV).  Changes in
the DFT gap energy, in turn, correspond to the excited-state forces
given by constrained DFT.  Our results therefore imply that the forces
are indeed given reliably by the constrained DFT theory, even though
the exciton binding energies are large and the absolute values of the
DFT band gaps are in poor agreement with the excitonic
energies.\cite{note:singlet-triplet_relaxations}

The singlet excitation energy for the ground state structure of the
single chain (3.41 eV) is in good agreement with experimental
absorption spectra which show maximum absorption around 3.5
eV.\cite{grem_1992,grem_1993} We find that the singlet-triplet
splitting depends rather weakly on the relaxations, changing from 0.79
eV at $x$=0 to 0.57 eV at $x$=1. We therefore expect the splitting for
the relaxed excited state structure to be in the region of 0.5 eV,
which is consistent with the splitting of 0.7-0.8 eV obtained from
emission spectra of PPP derivatives.\cite{hertel_2001}

We have analyzed the reasons for the success of the constrained DFT
approach for calculating excited state geometries.  The details of
this analysis will be published elsewhere, but we conclude that the
various approximations should work well for the chain or solid when
the excitons are delocalized over many atoms and when the relaxations
are relatively small and do not include reconstructive changes such as
bond breaking.  The constrained DFT approach would be inaccurate if
the excitonic wave function contained contributions from more than one
electron or hole band, but this mixing could be calculated from BSE
calculations on high symmetry structures and then used in DFT studies
of large structures.  Our study suggests that the constrained DFT
approach is likely to work for other conjugated polymers.

Our main calculations are for single chains, in which the
electron-hole interaction is larger than in solids.  Therefore our
calculations are an even stronger test of the idea of neglecting the
electron-hole interaction when calculating the excited state
relaxations than would be encountered in the more technologically
relevant solid state.  In the solid, many chains are packed together
at van der Waals distances, which gives rise to three main additional
features: (i) electronic overlap between chains, (ii) van der Waals
interactions, and, in particular, (iii) the effect of inter-chain
dielectric screening on the electron-hole interaction.  The first of
these is well described within DFT while MBPT includes the second and
third.

To study the effects of inter-chain dielectric screening we considered
a 3D crystalline array of PPP chains in which the electronic overlap
between chains is small and consequently the DFT results are only
weakly perturbed.  The calculated QP correction to the DFT gap and the
electron-hole interaction are, however, both reduced by more than 1~eV
as a consequence of the more effective dielectric screening in a 3D
solid.  The QP state, which describes an additional electron or hole
on one chain, shows significant interaction with the polarizable
neighboring chains, and the QP gap closes.  In addition, the QP
corrections also depend on $x$, indicating that the intra-chain
screening changes when the geometry relaxes.  The exciton energies
are, however, only weakly affected by the changes in the inter- or
intra-chain screening.  In the solid the singlet (triplet) energy
ranges from 3.66 eV (3.14 eV) at $x$=0 to 2.55 eV (2.23 eV) at $x$=1,
which is only a little higher than the single-chain results.  This
insensitivity derives from the charge-neutral character of the
exciton, which is much less influenced by electrostatic screening than
charged single-particle excitations.  Similar effects have also been
observed for other polymers by van der Horst \textit{et
al}.\cite{horst} The changes in the excitation energies as $x$ varies
from 0 to 1 (--1.11 eV for the singlet, --0.91 eV for the triplet) are
very similar to the change of the DFT gap (--0.95 eV), confirming the
reliability of the constrained-DFT forces.  The weak interactions
between chains have to be handled with care, however, and the
self-trapping of the exciton in the solid state will be the subject of
a further study.

In summary, we have shown that while the electron-hole interaction is
very important in determining the excitonic energies in PPP it is less
important for the excited state geometries.  We have demonstrated that
a simple constrained DFT approach is sufficient for calculating
relaxations in the low-energy excited states of PPP.  In this case we
find that the structural relaxations extend over about 8 monomers,
leading to an energy reduction of 0.22 eV and a Stokes shift of 0.40
eV.  The DFT method is tractable for studying the influence of
structural disorder on the optical properties of polymers.

We thank Neil Greenham for discussions.  We acknowledge financial
support from the Engineering and Physical Sciences Research Council of
the United Kingdom and the Deutsche Forschungsgemeinschaft.  PDH
acknowledges the support of Sidney Sussex College, Cambridge.


\begin{thebibliography}{}
\bibitem{footnote_gaps} The Stokes shift can also be calculated as the
difference between the constrained DFT total energy for the excited
state and the DFT energy for the ground state, but this gives an
almost identical result.
\bibitem{note:singlet-triplet_relaxations} These results also imply
that the structural relaxations and Stokes shifts in the lowest
singlet and triplet states of PPP should be very similar.
\end{thebibliography}
\end{document}